\documentclass[10pt,conference]{IEEEtran}

\usepackage{graphics} 
\usepackage[pdftex]{graphicx}
\usepackage{times} 
\usepackage [cmex10]{amsmath} 
\usepackage{amssymb}  
 \IEEEoverridecommandlockouts
\title{Capacity of  Fading Gaussian Channel with an Energy Harvesting Sensor Node
\thanks{This work is partially supported by a grant from ANRC to Prof. Sharma.}
\thanks{This work was done when Prof. Viswanath was visiting Indian Institute of Science. The visit of Prof. Viswanath is supported by DRDO-IISc Programme on Advanced Mathematical Engineering.}}
\author{\IEEEauthorblockN{R Rajesh}
\IEEEauthorblockA{CABS, DRDO\\
Bangalore, India\\
Email: rajesh81r@gmail.com}
\and
\IEEEauthorblockN{Vinod Sharma }
\IEEEauthorblockA{Dept. of ECE\\
Indian Institute of Science\\
Bangalore, India\\
Email: vinod@ece.iisc.ernet.in }\and
\IEEEauthorblockN{Pramod Viswanath}
\IEEEauthorblockA{Electrical and Computer Engineering Dept.\\
University of Illinois\\
Urbana-Champaign, USA\\
Email: pramodv@uiuc.edu}}
\vspace{0.01cm}

\begin{document}
\maketitle
\thispagestyle{empty}
\pagestyle{empty}
\begin{abstract}
Network life time maximization is becoming an important design goal in wireless sensor networks. Energy harvesting has recently become a preferred choice for achieving this goal as it provides near perpetual operation. We study  such a sensor node with an energy harvesting source and  compare various architectures by  which the harvested energy is used.  We find its Shannon capacity when it is transmitting its observations over a fading  AWGN channel with perfect/no channel state information provided at the transmitter.  We  obtain an achievable rate when there are inefficiencies in energy storage and the capacity when energy is spent in activities other than transmission.
\end{abstract}

\section{Introduction}
\label{intro} 
Sensor nodes  deployed for monitoring a random field are characterized by  limited battery power, computational resources  and storage space. Often, once deployed the battery of these nodes are not changed.  Hence when the battery of a node is exhausted, the node dies. When sufficient number of nodes die, the network may not be able to perform its designated task. Thus the life time of a network is an important characteristic of a sensor network (\cite{bhardwaj}) and it depends on the life time of a node. 

Recently, energy harvesting techniques (\cite{kansal}, \cite{niyato}) are gaining popularity for increasing the network life time. Energy harvester harnesses energy from environment or other energy sources (e.g., body heat) and converts them to electrical energy. Common energy harvesting devices are solar cells, wind turbines and  piezo-electric cells, which extract energy from the environment. Among these,  harvesting solar energy through photo-voltaic effect seems to have emerged as a technology of choice for many sensor nodes (\cite{niyato}, \cite{raghunathan}). Unlike for a battery operated sensor node, now there is potentially an \textit{infinite} amount of energy available to the node. However, the source of energy and the energy harvesting device may be such that the energy cannot be generated at all times (e.g., a solar cell).  Furthermore the rate of generation of energy can be limited. Thus the new design criteria may be to match the energy generation profile of the harvesting source with the energy consumption profile of the sensor node. If the energy can be \textit{stored} in the sensor node then this matching can be considerably simplified. But the energy storage device may have limited capacity. The energy consumption policy  should be designed in such a way that the node can perform satisfactorily for a long time. In \cite{kansal} such an energy/power management scheme is called  \textit{energy neutral operation}. 

We study the Shannon capacity of such an energy harvesting sensor node transmitting over a fading Additive White Gaussian Noise (AWGN) Channel. We provide the capacity under various energy buffer constraints and perfect/no channel state information at the transmitter (CSIT). We show that the capacity achieving policies are related to the throughput optimal policies (\cite{vinod1}). We also provide an achievable rate for this system with  inefficiencies in the energy storage. Finally we provide the capacity when energy is used not only in transmission but also for sensing, processing etc.

In the following we survey the relevant literature. Energy harvesting in sensor networks are studied in  \cite{kansal1} and \cite{rahimi}. Conditions for energy neutral operation for various models of energy generation and consumption are provided in \cite{kansal}. A practical solar energy harvesting sensor node prototype is described in \cite{jiang}.  In \cite{jaggi} the authors study  optimal sleep-wake cycles for  event detection probability in sensor networks. 

 Energy harvesting architectures can be  often divided into two major classes. In {\it{Harvest-use}} (HU), the harvesting system directly powers the sensor node and when sufficient energy is not available the node is disabled. In {\it{Harvest-Store-Use}} (HSU) there is a storage component that stores the harvested energy and also powers the sensor node. The storage can be single or double staged (\cite{kansal},~\cite{jiang}).

Various throughput and delay optimal energy management policies for energy harvesting sensor nodes are provided in \cite{vinod1}.  These energy management policies in \cite{vinod1} are extended in  various directions in \cite{vinod2} and \cite{vinod3}. For example, \cite{vinod2} provides some efficient MAC policies for energy harvesting nodes. In \cite{vinod3} optimal sleep-wake policies are obtained for such nodes. Furthermore, \cite{vinod4} considers jointly optimal routing, scheduling and power control policies for networks of energy harvesting nodes.

In a recent contribution, optimal energy allocation policies over a finite horizon and  fading channels are studied  in \cite{sg}. An information theoretic analysis of an energy harvesting communication system is provided in \cite{uluk}.

The capacity of a fading Gaussian channel with channel state information (CSI) at the transmitter and receiver and at the receiver alone are provided in \cite{varia}. It was shown that optimal power adaptation when CSI is available both at the transmitter and the receiver is `water filling' in time. An excellent survey on fading channels is provided in  \cite{proakis}.

We consider the problem of determining the information-theoretic capacity of an energy harvesting sensor node transmitting its observation over a fading AWGN channel. We  provide an achievable rate when there are  inefficiencies in energy storage. We also compute the capacity when energy is spent in activities other than transmission. We address the case without fading in \cite{ncc11}. Our results can be useful in the context of green communication (\cite{green1},~\cite{green2}) when solar and/or wind energy can be used by a base station (\cite{wind}).


The paper is organized as follows. Section \ref{model} describes the system model. 
Section \ref{stability} provides the capacity for a single node under idealistic assumptions.
 We show that the capacity achieving policy is related to  throughput-optimal policy.
 Section \ref{opt} obtains an achievable rate with inefficiencies in the energy storage system.  
 Section \ref{alloc} provides the capacity when energy is consumed in activities other than transmission. Section \ref{conclude} concludes the paper.

\section{Model and notation}
\label{model} In this section we present our model for a single energy harvesting sensor node.

\begin{figure}[h]
\begin{center}
\includegraphics[height=1.3in, width=3.5in]{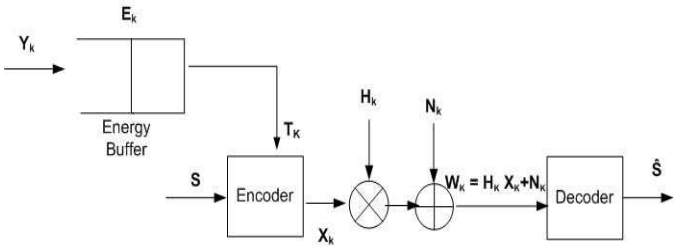}
\caption{The model} \label{fig1}
\end{center}
\end{figure}
We consider a sensor node (Fig. \ref{fig1}) which is sensing and generating data to be transmitted to a central node via a discrete time AWGN Channel.  We assume that transmission consumes most of the energy in a sensor node and ignore other causes of energy consumption (this is true for many low quality, low rate sensor nodes (\cite{raghunathan})).  The sensor node is able to replenish energy by $Y_k$  at time $k$ via an energy harvesting source. The energy available at the node at time $k$ is $E_k$. This energy is stored in an energy buffer with an infinite capacity. In  later part of the paper we will relax some of the assumptions made in this section.

The node uses energy $T_k$ at time $k$ which depends on $E_k$ and $T_k \le E_k$. The process $ \{ E_k \}$ satisfies  
\begin{eqnarray}
E_{k+1}  = (E_k - T_k) + Y_k. \label{eqn2}
\end{eqnarray} 

We  assume that  $\{ Y_k \}$  is  stationary, ergodic. This assumption is general enough to cover most of the stochastic models developed for energy harvesting.  Often the energy harvesting process has time varying  statistics (e.g., solar cell energy harvesting  depends on the time of the day). Such a process can  be approximated by piecewise stationary processes. As in \cite {vinod1}, we can indeed consider $\{Y_k\}$ to be periodic, stationary, ergodic.

The encoder receives a message $S$ from the node and generates an $n$-length codeword to be transmitted on the fading AWGN channel. We assume flat, fast,  fading. At time $k$ the channel gain is $H_k$ and takes values in $\mathcal{H}$. For simplicity we assume $\mathcal{H}$ to be countable. It can be easily extended to the case with set of real numbers $\mathcal{H}$. The sequence $\{ H_k \}$ is assumed independent identically distributed $(iid)$, independent of the  energy generation sequence $\{ Y_k \}$. The channel output at time $k$  is $W_k=H_k X_k+N_k$ where $X_k$ is the channel input at time $k$ and  $\{N_k\}$ is  \emph{iid} Gaussian noise with zero mean and variance $\sigma^2$. The decoder receives $Y^n \stackrel{\Delta}{=}(Y_1,...,Y_n)$ and reconstructs  $S$ such that the probability of decoding error is minimized. Also, the decoder has perfect knowledge of the channel state $H_k$.
 
We  obtain the information-theoretic capacity of this channel. This of course assumes that there is always data to be sent at the sensor node. This channel is essentially different from the usually studied systems in the sense that the transmit power and coding scheme depend on the energy available in the energy buffer at that time. 




\section{Capacity for the Ideal System}
 \label{stability} 
In this section we obtain the capacity of the channel with an energy harvesting node under ideal conditions : infinite energy buffer, energy  consumed in transmission only,  no inefficiencies in storage and perfect CSI at the transmitter.  Several of these assumptions will be removed in later sections. We always assume that the receiver has perfect CSI.

The system starts at time $k=0$ with an empty energy buffer and $E_k$ evolves with time depending on $Y_k$ and $T_k$. Thus $\{E_k,~k \ge 0\}$ is not stationary and hence $\{T_k\}$ may also not be stationary.  In this setup, a reasonable general assumption is to expect $\{T_k\}$ to be asymptotically stationary. One can further generalize it to be Asymptotically Mean Stationary (AMS), i.e.,
\begin{equation}
\lim_{n \to \infty} \frac{1}{n} \sum_{k=1}^n P[T_k \in A]= \overline{P}(A)
\end{equation}
exists for all measurable $A$. In that case $\overline{P}$ is also a probability measure and is called the \emph{stationary mean} of the AMS sequence (\cite{gray}).

If the channel input $\{X_k\}$ is AMS, then it can be easily shown that for the fading AWGN channel $\{(X_k,W_k),~k \ge 0\}$ is also AMS. Thus the channel capacity of our system is (\cite{gray})
\begin{equation}
\label {eqnn1}
C= \sup_{p_x} \overline{I}(X;W)= \sup_{p_x} ~ \limsup_{n \to \infty}\frac{1}{n} I(X^n,W^n)
\end{equation}
where under $p_x$, $\{X_n\}$ is an AMS sequence, $X^n=(X_1, ..., X_n)$ and the supremum is over all possible AMS sequences $\{X_n\}$. For a fading  AWGN channel capacity achieving $X_k$ is zero mean Gaussian with variance $T_k$ where $T_k$ depends on  the power control policy used and is assumed AMS. Then  $E[T] \le E[Y]$ where $E[T]$ is the mean of $T$ under its stationary mean. If $R< C$ then one can find a sequence of codeword's with code length $n$ and rate $R$ such that  the average probability of error goes to zero as $n \to \infty$ (\cite{gray}).

In the following we obtain $C$ in \eqref{eqnn1} for our system. 

We need the following definition. 

\emph{Pinskar Information Rate} (\cite{gray}): Given an AMS  random process $\{(X_n, W_n)\}$ with standard alphabets (Borel subsets of Polish spaces) $A_X$ and $A_W$, the Pinskar information rate is defined as $I^{*} (X;W)= \sup_{q,r} \overline{I} (q(X); r(W))$ where the supremum is over all quantizers $q$ of 
$A_X$  and $r$ of $A_W$.

 It is known that, $ I^*(X;W) \le \overline{I}(X;W)$. The two are equal if the alphabets are finite. The capacity in \eqref{eqnn1} involve 
 $ \overline{I}(X,W)$. But $I^{*} (X;W)$ is easier to handle.

We also need the following Lemma for proving the achievability of the capacity of the channel. This Lemma holds for $I^*$ but not for $\overline{I}$.

{\bf Lemma 1} (\cite{gray} Lemma  6.2.2):  Let  $\{X_n, W_n\}$  be  AMS  with distribution $P$ and stationary mean $\overline{P}$. Then $I^*_P(X;W) = I^*_{\overline{P}}(X;W)$.

{\bf Theorem 1} For the energy harvesting system with perfect CSIT  $C= 0.5~E_{H}[\log(1+\frac{H^2T^*(H)}{\sigma^2})]$ where

\begin{equation}
T^* (H) = \left(\frac{1}{H_0}-\frac{1}{H}\right)^+,\label{fto} 
\end{equation}
and $H_0$ is chosen such that $E_H[T^*(H)] = E[Y]$.

{\bf Proof:}  \emph{Achievability}: Let $T_k'=T^*(H_k)$ with $T^*$ defined in \eqref{fto} with $E[T^{*}(H)]=E[Y]-\epsilon$ where $\epsilon > 0$ is a small constant. Since $\{H_k\}$ is $iid$, $\{T_k'\}$ is also $iid$. We take $T_k= min(E_k,T_k')$. Then from \cite{un}, $E_k \to \infty~a.s.$ Therefore, as $T^*(H)$ is upper bounded, $\lim_{n \to \infty} \sup_{k \ge n} |T_k-T^*(H_k)| \to 0~a.s.$

 Let $\{X_k^{'}\}$ be $iid$ Gaussian with mean zero and variance one. The channel codeword $X_k= sgn(X_k')\min(\sqrt{T_k} |X_k'|,\sqrt{E_k})$ where $sgn(x)=1$, if $x \ge 0$ and $-1$ otherwise.  This is an AMS sequence with the stationary mean being the distribution of  $\sqrt{T^*(H_k)}X_k'$. Then since AWGN channel under consideration is AMS ergodic (\cite{gray}), $(X,W)\stackrel{\Delta}{=} \{(X_k,W_k), ~k \ge 1\}$ is AMS ergodic.   

By using Lemma 1, $I^*(X;W)=$ $\sup_{q,r}$ $\limsup_{k \to \infty}$ ${1}/{k} ~I(q(X^k);r(W^k))$ $=I^*(\hat{X},\hat{W})$ where $I^*(\hat{X},\hat{W})$ corresponds to the limiting $iid$ sequence $\{\hat{X}_k,\hat{W}_k\}$ with  $\hat{X}_k  = T_k'X_k'$ and $\hat{W}_k$ is the corresponding channel output.

Also, since the mutual information between two random variables is the limit of the  mutual information between their quantized versions \cite{Cover04elements}, $I^*(\hat{X},\hat{W})=I(\hat{X},\hat{W})=0.5~E_H [\log(1+{H^2 T^{*}(H))}/\sigma^2)]$. We can show that as $\epsilon \to 0$,  $0.5 ~E_H [\log(1+{H^2 T^{*}(H))}/\sigma^2)] \to C$ defined in the statement of the theorem.

\emph {Converse  Part}: Let there be a sequence of codebooks for our system with rate $R$ and average probability of error going to 0 as $n \to \infty$. If $\{X_k(S),~k=1,...,n\}$ is a codeword for message $S \in \{1,...,2^{nR}\}$ then  ${1}/{n} \sum_{k=1}^n X_k(S)^2 \le 1/n \sum_{k=1}^n Y_k \le E[Y] + \delta$ for any $\delta >0$  with a large probability for all $n$ large enough. 
Hence by the converse in the fading AWGN channel case (\cite{varia}),
 $\limsup_{k \to \infty} {1}/{k} I(X^k;W^k)\le  0.5~E_H[\log(1+{H^2 T^*(H))}/\sigma^2)]$ for $T^*(H)$ given in \eqref{fto}.

Combining the direct part and converse part completes the proof. ~~~~~~~~~~~~~~~~~~~~~~~~~~~~~~~~~~~~~~~~~~~~~~~~~~~~~~~~~~ {\raggedleft{$\blacksquare$}}
\vspace{0.1cm}

Thus we see that the capacity of this fading channel is same as that of a node with average power constraint $E[Y]$ and instantaneous power allocated according to  `water filling' power allocation, i.e., the hard energy constraint of $E_k$ at time $k$ does not affect its capacity.
The capacity achieving signaling for our system is  $X_k=sgn(X_k')\min(\sqrt{T_k} |X_k'|,\sqrt{E_k})$, where $\{X_k'\}$ is $iid$ $\mathcal{N}(0,1)$ and $T^*(H)$ is defined in \eqref{fto}. 

When  no CSI is available at the transmitter, take  $X_k=sgn(X_k')\min( |X_k'|,\sqrt{E_k})$ where $\{X_k'\}$ is $iid$ $\mathcal{N}(0,E[Y])$ and as in Theorem 1 this approaches the capacity of  $0.5~E_H[\log(1+{H^2 E[Y]}/{\sigma^2})]$. 

In \cite{vinod1}, a system with   a data buffer at the node which stores data sensed by the node \emph{before} transmitting it over the fading AWGN channel, 
is considered.
 The stability region (for the data buffer) for the 'no energy-buffer' and 'infinite energy-buffer'
 corresponding to the harvest-use and harvest-store-use architectures with perfect/no CSIT are provided. 
The throughput optimal policies in \cite{vinod1} are  related to the 
 Shannon capacity achieving energy management policies provided here for  the infinite buffer case.
Also the capacity is the same as the maximum throughput obtained  in the data-buffer case in \cite{vinod1}.

If there is no energy buffer to store the harvested energy then at time $k$ only $Y_k$ energy is available. Thus $X_k$ is peak power limited to $Y_k$. The capacity achieving distribution for an AWGN channel with peak power constraint $Y_k=y$ has been studied (\cite{peak1}, \cite{peak2}, \cite{peak3}) and is not Gaussian. Let $X(y,\sigma^2)$ be a random variable with the capacity achieving distribution for AWGN channel with peak power constraint $y$ and noise variance $\sigma^2$. In general this distribution is discrete. Thus, if CSIT is exact then the transmitter will transmit $X(y,\sigma^2/h^2)$ at time $k$ when $Y_k=y$ and $H_k=h$. Therefore the ergodic capacity is $0.5 E_{YH}[I(X(Y,\sigma^2/H^2);W)]$. If there is no CSIT then we can transmit $X(y,\sigma^2)$ and the  corresponding capacity is   $0.5 E_{YH}[I(X(Y,\sigma^2);W)]$.

Our basic model in Fig. 1 considers the case when the battery charge changes in every channel use. If the channel rate is high it is possible to think of time index $k$, as $k-th$ slot consisting of $N$ channel uses. Then energy $E_k$ is available for $N$ channel uses during each slot. If the channel gain changes in $iid$ fashion per channel use, the capacity stated in Theorem 1 holds per channel use. Here $E[Y]$ denotes the mean energy harvested per channel use.

\section{Achievable Rate with Energy Inefficiencies}
\label{opt}
In this section we make our model more realistic by taking into account the inefficiency in storing energy in the energy buffer  and the leakage from the energy buffer (\cite{jiang}) for HSU architecture.

We assume that if energy $Y_k$ is harvested at time $k$, then only energy $\beta_1Y_k$ is stored in the buffer  and  energy $\beta_2$ gets leaked in each slot where $0 \le \beta_1 \le 1$ and $0 \le \beta_2<\infty$.   Then \eqref{eqn2} becomes
\begin{eqnarray}
E_{k+1}  = ((E_k - T_k)-\beta_2)^{+} + \beta_1Y_k. \label{eqne1}
\end{eqnarray} 

The energy can be stored in a super capacitor and/or in a battery.
 For a supercapacitor, $\beta_1 \ge 0.95$ and for the Ni-MH battery (the most commonly used battery)
 $\beta_1 \sim 0.7$. The leakage  $\beta_2$ for the super-capacitor as well as the battery is close to 0 but for the super capacitor it may be somewhat larger.

In this case, similar to the achievability of Theorem 1 we can show that the following rates are achievable in the no CSIT and Perfect CSIT  case respectively
\begin{eqnarray}
 R_{S-NCSIT} = 0.5~E_H\left[\log\left(1+\frac{H^2(\beta_1 E[Y]-\beta_2)}{\sigma^2}\right)\right], \label{eqnr1}\\
R_{S-CSIT} = 0.5~E_H\left[\log\left(1+\frac{H^2(\beta_1 T(H)-\beta_2)}{\sigma^2}\right)\right],\label{eqnr2}
 \end{eqnarray}
where $T(H)$ is a power allocation policy such that \eqref{eqnr2} is maximized subject to $E_H[T(H)]\le E[Y]$
This policy is neither capacity achieving nor throughput optimal \cite{vinod1}.

An achievable rate when there is no buffer and perfect CSIT is
 \begin{eqnarray}
C= E_{YH}[I(X(Y,H);W)],\label{eqnr3}
 \end{eqnarray}
where $X(y,h)$ is the distribution that maximizes the capacity subject to peak power constraint $y$ and fade state $h$. It is also shown in \cite{peak3} that for $\sqrt{y} <1.05$, the capacity has a closed form expression 

\begin{eqnarray}
C(y)= y-\int_{-\infty}^{\infty} \frac{e^{-x^2/2} \log {cosh(y-\sqrt{y}x)}}{\sqrt{2\pi}} dx. \label{eqnrr3}
\end{eqnarray}

When there is no buffer and no CSIT the distribution that maximizes the capacity cannot be chosen as in \eqref{eqnr3} and the capacity is less than the capacity given in \eqref{eqnr3}. The capacity in  \eqref{eqnr3}  is without using buffer and hence $\beta_1$ and $\beta_2$ do not affect the capacity. Hence unlike in  Section III, \eqref{eqnr3} may be larger than \eqref{eqnr1} and \eqref{eqnr2} for certain range of parameter values. We will illustrate this by an example.
 
 Another achievable policy for the system with  an energy buffer  
 with storage inefficiencies is to use the harvested energy $Y_k$ immediately instead of storing in the buffer. 
The remaining energy after transmission is stored in the buffer. We call this \emph{Harvest-Use-Store} (HUS) architecture.
 For this case, \eqref{eqne1} becomes
 \begin{eqnarray}
E_{k+1}  = ((E_k + \beta_1(Y_k-T_k)^{+}- (T_k-Y_k)^{+})^{+}-\beta_2)^{+}. \label{eqne2}
\end{eqnarray} 
Find the largest constant $c$ such that $ \beta_1 E[(Y_k-c)^{+}] \ge  E[(c-Y_k)^{+}]+ \beta_2$. Of course $c<E[Y]$. When there is no CSIT, this is the largest $c$ such that taking $T_k= \min(c-\delta,E_k)$, where $\delta >0$ is any small constant,  will make $E_k \to \infty~a.s.$ and hence $T_k \to c~a.s.$ Then, as in Theorem 1, we can show that
 \begin{eqnarray}
 R_{US-NCSIT} = 0.5~E_H\left[\log\left(1+\frac{H^2c}{\sigma^2}\right)\right] \label{eqnc1}
 \end{eqnarray}
 is an achievable rate.
 
 When there is perfect CSIT, 'water filling' power allocation can be done subject to average power constraint of $c$ and the achievable rate is
  \begin{eqnarray}
 R_{US-CSIT} = 0.5~E_H\left[\log\left(1+\frac{H^2T^{*}(H)}{\sigma^2}\right)\right] \label{eqnc2}
 \end{eqnarray}
  where $T^{*}(H)$ is the 'water filling' power allocation with $E[T^{*}(H)] = c$.

Equation \eqref{eqne1} approximates the system 
where we have only rechargeable battery while \eqref{eqne2} 
approximates the system where the harvested energy is first stored in a super capacitor
 and after initial use transferred to the battery.

We illustrate the achievable rates mentioned above via an example.

\subsection{Example 1}

\begin{figure}[!h]
\centering
\includegraphics [height=3 in, width=3.5in ]{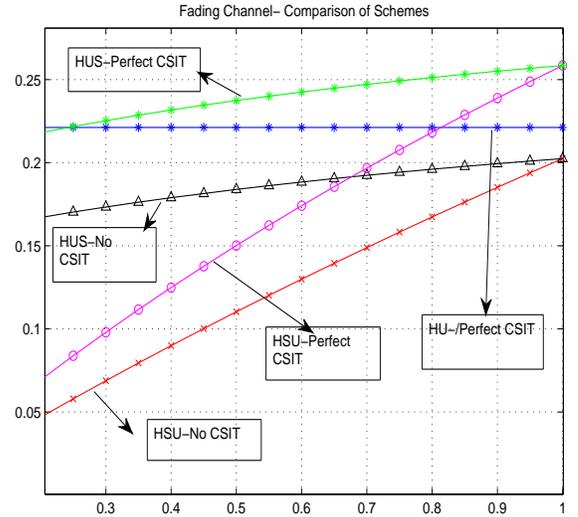}
\caption{Rates for various architectures}
\label{figineff}
\end{figure}

Let the process $\{Y_k\}$   be $iid$ taking values in  $\{0.5,~ 1\}$ with probability $\{0.6,~  0.4\}$ . We take the
 loss due to leakage $\beta_2=0$. The fade states are $iid$ taking values in $\{0.4, ~0.8,~1\}$ with probability $\{0.4,~ 0.5,~ 0.1\}$. In Figure \ref{figineff} we compare the various architectures discussed in this section for varying storage efficiency $\beta_1$. The capacity for the no buffer case with perfect CSIT is computed using equations \eqref{eqnrr3} and \eqref{eqnr3}.

From the figure we observe the following

\begin{itemize}
	\item Unlike the ideal system, the $HSU$ (which uses infinite energy buffer) performs worse than the $HU$  (which uses no energy buffer) when storage efficiency is poor for the perfect CSIT case.
	\item When storage efficiency is high, $HU$ policy performs worse compared to $HSU$ and $HUS$ for perfect CSIT case.
	\item $HUS$ performs the best for No/Perfect CSIT compared to $HSU$.
	\item For $\beta=1$, the $HUS$ policy and $HSU$ policy are the same for both perfect CSIT and no CSIT.
	\item The availability of CSIT and storage architecture plays an important role in determining the achievable rates.
\end{itemize}

Thus if we judiciously use a combination of a super capacitor and a battery with perfect CSIT one may obtain a better performance.

 \section{ Capacity with Energy Consumption in Sensing and Processing}
\label{alloc}
In recent higher end sensor nodes, sensing, processing and receiving from other nodes also require significant energy apart from that used in transmission. 

We  assume that energy $Z_k$  is consumed by the node (if $E_k \ge Z_k$) for
 sensing and processing etc at time instant $k$. For transmission at time $k$, only $E_k-Z_k$ is available.
 $ \{ Z_k \}$ is assumed a stationary ergodic sequence. The rest of the system is as in Section \ref{model}. To bring out the effects of energy consumed on processing, sensing etc  explicitly, we will not consider the  storage inefficiencies.

First we extend the achievable policies in Section \ref{stability} to incorporate this case. When there is perfect CSIT, we use the signaling scheme   $X_k=sgn(X_k')\min(\sqrt{T_k} |X_k'|,\sqrt{E_k})$, where $\{X_k'\}$ is $iid$ $\mathcal{N}(0,1)$ and $T^*(H)$ is the optimum power allocation such that  $E[T^*(H)]= E[Y]-E[Z]-\epsilon$. When  no CSI is available at the transmitter, we use $X_k=sgn(X_k')\min( |X_k'|,\sqrt{E_k})$ where $\{X_k'\}$ is $iid$ $\mathcal{N}(0,E[Y]-E[Z]-\epsilon)$. The achievable rates are respectively,

\begin{gather}
R_{PE-CSIT}= 0.5~E_H\left[\log\left(1+\frac{H^2 T^*(H)}{\sigma^2}\right)\right],\label{abcd}\\
R_{PE-NCSIT}= 0.5~E_H\left[\log\left(1+\frac{H^2(E[Y]-E[Z]-\epsilon)}{\sigma^2}\right)\right].\label{abc}
\end{gather}

If the sensor node has two modes: Sleep and Awake then the achievable rates can be improved. The sleep mode is a power saving mode in which  the sensor only harvests energy and performs no other functions so that the energy consumption is minimal (which will be ignored). If $E_k < Z_k$ then we assume that the node will sleep at time $k$. But to optimize its transmission rate it can sleep at other times also. We consider a policy called \emph{randomized sleep policy} in \cite{vinod3},~ \cite{ncc11}. In  this policy at every time instant $k$ with $E_k \ge Z_k$ the sensor chooses to sleep with probability $p$ independent of all other random variables.

Let  $b(x)$ be the cost of transmitting $x$ and equals
\begin{eqnarray*}
b(x)= \begin{cases}
x^2+\alpha,~& \text{if $|x| >0$},\\
0,~& \text{if $|x| = 0$}.
\end{cases}
\end{eqnarray*}
and $\alpha= E[Z]$. Observe that if we follow a policy that unless the node transmits, it sleeps,
 then $b$ is the cost function.

{\bf Theorem 2} Let $\mathcal{P}(H)$ be the feasible power allocation policies such that for $P(H) \in \mathcal{P}(H)$, $E_H[P(H)] \le E[Y]$.  For the energy harvesting system with processing energy transmitting over a fading Gaussian channel,

\begin{equation}
C= \sup_{P(H) \in \mathcal{P}(H)} \sup_{p_x: E[b(X)] \le P(H)} E[I(X;W)] \label{impee}
\end{equation}
is the capacity for the system. Denote by $P^*(H)$  the capacity achieving power allocation.

{\bf Proof:}  : Fix the power allocation policy  $P^*(H)$. Under $P^*(H)$, the achievability of  $\sup_{p_x: E[b(X)] \le P^*(H)} I(X;W)$ is proved using  the techniques provided in Theorem 2 of \cite {ncc11} for the non-fading case. The achievability in \cite {ncc11} is proved by  showing that this rate can be achieved by an $iid$ signaling. Finally it is also shown that we can achieve this rate by a signaling scheme that satisfies the energy constraints. Using this along with finding the expectation w.r.t the optimum power allocation scheme completes the achievability proof.

The converse follows via Fano's inequality as in, for example, \cite{varia} for fading AWGN channel. In the converse, we use the fact that $C(.)$ is concave. ~~~~~~~~~~~~~~~~~~~~~~~~~~~~~~~~~~~~~~~~~~~~~~~~~~~~{\raggedleft{$\blacksquare$}}
\vspace{0.1cm}

It is interesting to compute the capacity \eqref{impee} and the capacity achieving distribution $p_X$ under state $H=h$. Let $P(h)$ be the power allocated in state $H$.  Without loss of generality, we can say that under $p_X$, with probability $p ~(0 \le p \le 1)$ the node sleeps and with probability $1-p$ it transmits with a density $f$. We can show using Kuhn-Tucker conditions that density $g$ of the received symbol when the node is transmitting with density $f$ under $H=h$ is
\begin{equation}
g(a)=\left(k_1e^{-a^2k_2}-\frac{p e^{-a^2/2\sigma^2}}{(1-p)\sqrt{2\pi\sigma^2}}\right)^{+},\label{newcorr}
\end{equation}
where $k_1$ and $k_2$ are positive constants such that the cost constraint $E[b(X)] \le P(h)$ is satisfied. We have found that at the optimal $p$ the term in \eqref{newcorr} is always non-negative ( thus $(~)^+$ is not needed at optimal $p$) which implies that when ever the node is awake, under $H=h$, the density $f$ is Gaussian with mean zero and variance $P(h)/(1-p)-E[Z]$. 

The optimal power allocation policy $P^*(H)$ that maximizes \eqref{impee} is  'water filling'.

\subsection{Example 2}

\begin{figure}[ht]
\centering
\includegraphics [height=3in, width=3.5in ]{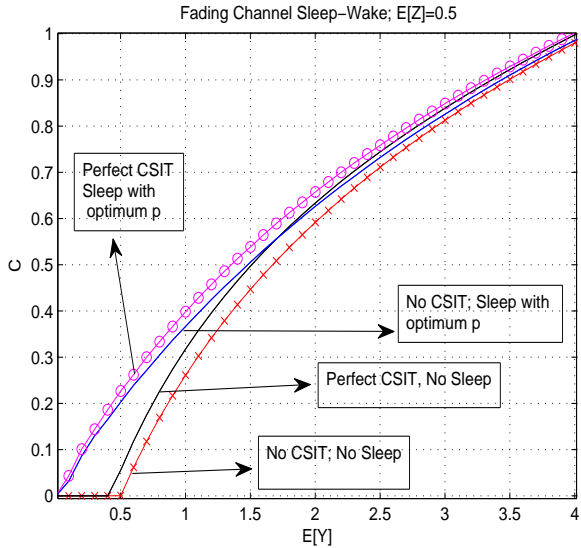}
\caption{Comparison of sleep wake policies}
\label{figsleep}
\end{figure}

Let the fade states take values in $\{0.5,~1,~1.2\}$ with probabilities $\{0.1,~0.7,~0.1\}$. We  take $\alpha=E[Z]=0.5,~ \sigma^2=1$. We compare the capacity for the cases with perfect and no CSIT when there is no sleep mode supported (Equation \eqref{abcd}, \eqref{abc}) and with the optimal sleep probability in Figure \ref{figsleep}.

From the figure we observe the following
 \begin{itemize}
	\item For small values of $E[Y]$, the availability of CSIT improves the rate significantly.
	\item The randomized sleep wake policy improves the rate significantly when $E[Y] < E[Z]$ or they are comparable.
	\item The sensor node chooses not to sleep when $E[Y] >> E[Z]$.
\end{itemize}

\section{Conclusions}
\label{conclude}
In this paper  the Shannon capacity of  an energy harvesting sensor node transmitting over a fading AWGN Channel is provided. It is shown that the  capacity achieving policies are related to the throughput optimal policies provided in \cite{vinod1} for infinite buffer case. Achievable rates are provided when there are inefficiencies in energy storage.  Also, the capacity is provided when the energy is consumed for activities other than transmission.

\bibliographystyle{IEEEtran}
\bibliography{mybibfilefade}

\end{document}